\documentclass[prl,aps,twocolumn]{revtex4}
\usepackage{bm,graphicx,amsmath} \usepackage{bbm}

\newcommand{\br}[1]{\langle #1|}
\newcommand{\ke}[1]{|#1\rangle}

\newcommand{\da}{^\dagger}

\newcommand{\pt}[1]{\left( #1 \right)}
\newcommand{\pq}[1]{\left[ #1 \right]}
\newcommand{\pg}[1]{\left\{ #1 \right\}}

\begin{document} \title{Cooling carbon nanotubes to the phononic
ground state with constant electron current} \author{Stefano
Zippilli,$^{1}$ Giovanna Morigi,$^{1}$ and Adrian Bachtold~$^{2}$}
\affiliation{ $^{1}$Departament de F\'{i}sica, Universitat
Aut\`{o}noma de Barcelona, E-08193 Bellaterra, Spain,\\
$^2$ CIN2(CSIC-ICN), Campus de la UAB, E-08193 Bellaterra, Spain.}
\date{\today}

\begin{abstract} We present a quantum theory of cooling of a
mechanical resonator using back-action with constant electron
current. The resonator device is based on a doubly clamped
nanotube, which mechanically vibrates and acts as a double quantum
dot for electron transport. Mechanical vibrations and electrons
are coupled electrostatically using an external gate. The
fundamental eigenmode is cooled by absorbing phonons when
electrons tunnel through the double quantum dot. We identify the
regimes in which ground state cooling can be achieved for
realistic experimental parameters. \end{abstract}

\maketitle

Cooling mechanical resonators has recently attracted considerable
interest, as it allows ultrasensitive detection of mass
\cite{Roukes, Zettl, Benjamin, Bockrath}, of mechanical
displacements \cite{Lahaye}, and of spin \cite{Rugar}. An
appealing prospect is to cool the mechanical resonator to its
phononic ground state. This achievement would open the possibility
to create and manipulate non classical states at the macroscopic
scale and to study the transition from the classical to the
quantum regime~\cite{Schwab-Roukes,SchwabEntanglement,Rabl04}.

The lowest phononic occupation number achieved so far has been
experimentally realized by cooling down the resonator in a
dilution fridge \cite{Naik}. Another promising approach is to
employ back-action, which consists of coupling mechanical
oscillations to visible or microwave photons
\cite{Kippenberg,Zeilinger,Heidmann,Bouwmeester,Lehnert,Kippenberg2,Marquardt,Genes,MartinPZ04}.
Recently, it has been theoretically
proposed~\cite{Martin,Blencowe,Clerk} and experimentally
demonstrated~\cite{Naik} that back-action cooling can be achieved
by coupling mechanical resonators to the constant electron current
through electronic nano-devices, such as normal-metal and
superconducting single-electron transistors. This approach is
appealing because it is easy to implement in a dilution fridge as
compared to techniques based on photons. Within this approach,
however, modest occupations of the phononic ground state have been
predicted~\cite{Martin,Blencowe,Clerk,MartinPZ04}. In particular,
using an analogy with laser cooling of atoms~\cite{Eschner},
back-action cooling by constant electron current in these systems
is essentially analogous to Doppler cooling~\cite{Blencowe}.

In this Letter, we theoretically demonstrate ground state cooling
of a mechanical nanotube resonator using constant electron
current. Specifically, the nanotube is employed both as the
mechanical resonator and the electronic device through which the
current flows. In addition, we consider the device layout in which
the nanotube acts as a double quantum dot (DQD). This setup allows
us to access an analogous regime of sideband cooling of the
oscillator~\cite{Eschner}. Calculations are carried out by
including the coupling of the resonator to the thermal noise of
the electrodes and the effect of electronic dephasing inside the
DQD. For realistic device parameters the temperature is lowered by
a factor of about 100. Moreover, we identify the regime in which
the oscillator ground state can reach more than 90\% occupation.

\begin{figure}[!th] \begin{center}
\includegraphics[width=8cm]{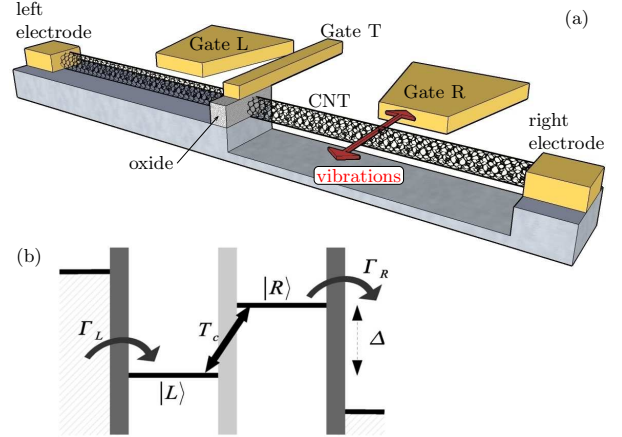} \caption{(a) Sketch
of the double quantum dot, carbon-nanotube (CNT) device. The dot
on the right is suspended, and mechanically oscillates. (b) The
single-electron levels $\ke{L}$ and $\ke{R}$, at (tunable)
frequency difference $\Delta$, are coherently coupled with
tunneling rate $T_c$. 
 }
\label{figlevels}
 \end{center}
  \end{figure}

The device layout is sketched in Fig.~\ref{figlevels}(a). The DQD
system is obtained by locally depleting a semiconducting nanotube
with gate T~\cite{Marcus,Kouwenhoven,Schonenberger,Lindelof}. The
dot on the right is suspended, so it can mechanically oscillate.
The nanotube is electrically contacted to two electrodes and the
electrochemical potential of the two dots is controlled with the
gates L and R. The DQD is voltage biased in order to have three
relevant quantum states of single-electron transport,
$\ke{0},\ke{L},\ke{R}$, see Fig.~\ref{figlevels}(b). Here,
$|L\rangle$ and $|R\rangle$ correspond to the excess electron
localized on the left and right dot with energy $\epsilon_L$ and
$\epsilon_R$, respectively. State $|0\rangle$, at zero energy,
corresponds to when both states are
unoccupied~\cite{Brandes05,Nazarov,KouwenhovenReview}. We denote
by $\Delta=(\epsilon_R-\epsilon_L)/\hbar$ the frequency difference
between $\ke{L}$ and $\ke{R}$. The coherent dynamics of the DQD is
given by the Hamiltonian \begin{eqnarray}\label{HDD} H_{\rm
DQD}=\hbar\pq{\epsilon_R \ke{R}\br{R}+\epsilon_L
\ke{L}\br{L}-T_c\pt{\ke{L}\br{R}+\ke{R}\br{L}}}, \end{eqnarray}
where $T_c$ is the tunneling rate between the dots.

The nanotube mechanical motion, considered here, is the
fundamental phononic mode, which is the flexural
eigenmode~\cite{Garcia-Sanchez,footnote:0}. We assume a high
quality factor $Q$ so 
the flexural mode is to good extent
uncoupled to the other phononic modes. We denote by $\omega$ the
oscillator frequency and by $a\da$ and $a$ the creation and
annihilation operators of a phonon at energy $\hbar\omega$. The
interaction between the mechanical motion and the conducting
electrons is obtained through the capacitive coupling between the
nanotube and gate R, which depends on their mutual distance. For
sufficiently small vibration displacements, the electron-phonon
interaction is described by Hamiltonian $H_{\rm
e-ph}=V\pt{a\da+a}$ where \begin{eqnarray}\label{V} V=\hbar \alpha
\ke{R}\br{R}, \end{eqnarray} and $\alpha$ is a coupling constant
which scales the mechanical effect~\cite{Clerk,Blencowe}. Cooling
is achieved by gating $\epsilon_L$ below $\epsilon_R$, so that a
phonon of the nano-mechanical oscillator is absorbed by electrons
tunneling from left to right.

We describe the cooling dynamics by using the master equation for
the density matrix $\rho$ of electron and mechanical degrees of
freedom, \begin{eqnarray} \dot{\rho}=-{\rm i}\omega\pq{a\da
a,\rho}-{\rm i}\pq{H_{e-ph},\rho}/\hbar+({\cal L}_{DQD}+{\cal
K})\rho \end{eqnarray} where \begin{eqnarray}{\cal L}_{\rm
DQD}\rho&=&-{\rm i}\pq{H_{\rm DQD},\rho}+
\Gamma_L/2\pt{2s_L\da\rho s_L-\rho
s_Ls_L\da-s_Ls_L\da\rho}\nonumber\\&+& \Gamma_R/2\pt{2s_R\rho
s_R\da-\rho s_R\da s_R-s_R\da s_R\rho}+\mathcal L_d\rho
\label{L_DD}\end{eqnarray} includes incoherent electron (tunnel)
pumping  at rate $\Gamma_L$ from the left electrode into state
$|L\rangle$, and (tunnel) extraction at rate $\Gamma_R$ from state
$|R\rangle$ into the right electrode, with $s_L=\ke{0}\br{L}$ and
$s_R=\ke{0}\br{R}$~\cite{footnote:1}. The term ${\cal
K}\rho=\gamma_-/2\pt{2a\rho a\da-a\da a\rho-\rho a\da
a}+\gamma_+/2 \pt{2a\da\rho a-aa\da\rho-\rho aa\da}$ describes the
thermalization of the flexural mode with the environment (other
phononic modes of the nanotube and of the electrodes) at the
cryogenic temperature $T$, where $\gamma_-=(\bar n_p+1)\gamma_p$,
$\gamma_+=\bar n_p\gamma_p$, with $\gamma_p=\omega/Q$ and $\bar
n_p=\pq{\exp\pt{\hbar\omega/k_BT}-1}^{-1}$~\cite{MartinPZ04,Wilson-Rae}.
Finally, $\mathcal L_d\rho$ is a process inducing incoherent
hopping between the two dots, which we will discuss later on.

In the limit $\gamma_p\ll\alpha\ll \omega$, a rate equation for
the population $p_n$ of the phononic energy level with $n$
excitations is derived, which reads $\dot
p_n=(n+1)A_-p_{n+1}-\pq{(n+1)A_++nA_-}p_n+nA_+p_{n-1}.$ Here, the
terms $A_+$ and $A_-$ give the rate with which the oscillator is
heated and cooled, respectively, by one phonon. The rate equation
describes the dynamics of a damped quantum oscillator when
$A_->A_+$, reaching the mean phonon number $\bar n_{\rm
St}=A_+/\gamma_{\rm tot}$ at the cooling rate $\gamma_{\rm
tot}=A_--A_+$. In this regime, the equation describes
thermalization with an effective bath at temperature $T_{\rm
osc}$, which 
can be well below the cryogenic temperature.

Terms $A_\pm$ are the sum of two contributions, the thermalization
rate via the environment and the coupling to the electronic
current. They read~\cite{Cooling,Zippilli} \begin{eqnarray}\label{A}
A_{\pm}=\frac{\gamma_\pm}{2}+\frac{2}{\hbar^2}{\rm Re}\pq{{\rm
Tr}\pg{V\pt{\mathcal L_{\rm DQD}\mp{\rm i}\omega}^{-1}V\rho_{\rm
St}} } \end{eqnarray} where the second term is the autocorrelation
function of the operator $V$, Eq.~(\ref{V}), taken over the
density matrix $\rho_{\rm St}$, which satisfies equation ${\cal
L}_{\rm DQD}\rho_{\rm St}=0$. The corresponding steady state
occupation number reads \begin{eqnarray}\label{n}
\bar n_{\rm St}= 
\frac{\gamma_p\bar n_p+\gamma_0\bar n_0 }{\gamma_p+\gamma_0}
\end{eqnarray} where $\gamma_0$ and $\bar n_0$ are the cooling
rate and the steady state phonon population solely due to
electron-phonon interaction ($\gamma_{\rm
tot}=\gamma_0+\gamma_p$). Equation~(\ref{n}) highlights the
competition between the active cooling process due to electron
transport and the thermalization with the surrounding environment,
showing that cooling is more effective when $\gamma_0\gg\gamma_p$.

We emphasize that the derived rate equation describes cooling of
the oscillator deep in the quantum regime. Moreover, note that,
using relation $k_BT_{\rm osc}=\hbar\omega\left(\ln\frac{\bar
n_{\rm St}+1}{\bar n_{\rm St}}\right)^{-1}$ and taking the
semiclassical limit of Eq.~(\ref{n}), we recover the formal
expression for the temperature obtained when the oscillator is
cooled by coupling to a single-electron transistor~\cite{Blencowe,Clerk}. However, while in the single-electron transistor setup
back-action cooling with constant electron current is essentially
analogous of Doppler cooling~\cite{Blencowe}, we access in our
case the sideband cooling regime and may hence reach unit
occupation of the ground state.

\begin{figure}[!t] \begin{center}
\includegraphics[width=8cm]{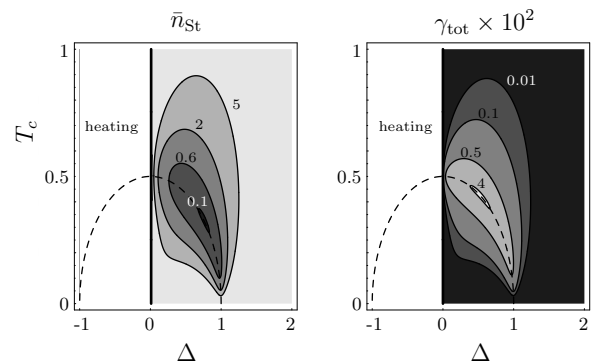}
\caption{Contour plot for average phonon number at steady state
$\bar n_{\rm St}$ and cooling rate $\gamma_{\rm tot}$ as a
function of $\Delta$ and $T_c$. Rates and frequencies are in units
of $\omega$. The other parameters are $\Gamma_L=10\omega,
\Gamma_R=0.1\omega, \alpha=0.1\omega$, $\gamma_p=10^{-4}\omega$
and $\bar n_p=24$.  The dashed lines indicates
the curve $\Delta^2+4T_c^2=\omega^2$. In the "heating" region one
finds $\bar n_{\rm St}> \bar n_p$.}\label{result} \end{center}
  \end{figure}
Figure~\ref{result} displays the contour plots of $\bar n_{\rm
St}$ and cooling rate $\gamma_{\rm tot}$ as a function of the
tunneling rate $T_c$ and of the frequency difference $\Delta$, for
$\Gamma_R\ll\omega$ and $\mathcal L_d=0$. For the chosen
parameters the lowest occupation is $\bar n_{\rm St}\sim 0.1$ and
the corresponding cooling rate is $\gamma_{\rm tot}\sim
0.03\omega$. To provide some typical numbers, for $T_c=2\pi\times
30$ MHz, $\Delta=2\pi\times 80 $ MHz, $\Gamma_L= 2\pi\times 1$
GHz, $\Gamma_R=2\pi\times 10$ MHz, and $\alpha=2\pi\times 10$
MHz~\cite{ExpParameter}, we find that the fundamental mode of a
nanotube resonator at $\omega=2\pi\times 100$ MHz and quality
factor $Q=10^4$ can be cooled from $T=120 $ mK ($\bar n_p=24$) to
$T_{\rm osc}=2$~mK ($\bar n_{\rm St}=0.1$) with cooling rate
$\gamma_{\rm tot}\sim 2\pi \times 3$ MHz. Note that if the initial
cryostat temperature ($\bar n_p$) is lower, the value of $\bar
n_{\rm St}$ will be smaller, see Eq.~(\ref{n}).

\begin{figure}[!t] \begin{center}
\includegraphics[width=8.5cm]{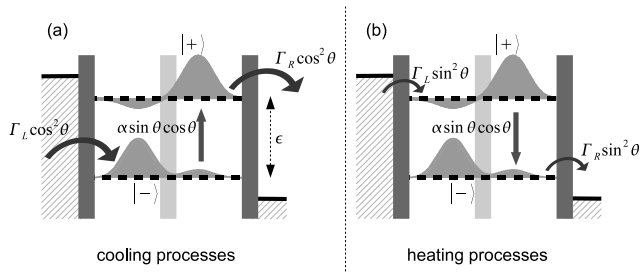}
\caption{Dominant cooling (a) and heating (b) processes in the
parameter region of optimal cooling in Fig.~\protect\ref{result}.
States $\ke{-}$ and $\ke{+}$ denote the bonding and antibonding
states at frequency difference $\epsilon=\sqrt{\Delta^2+4T_c^2}$.
The rates are reported in the new basis, with
$\tan\theta=2T_c/(\Delta+\epsilon)$. For $\Delta>0$, then
$\tan^2\theta<1$ and the processes in (a) are faster, so that the
resonator is cooled. }\label{processes}
 \end{center}
\end{figure} 
We remark that cooling is most efficient for tunneling rates
$T_c\sim \omega$ and for $\Delta\sim\sqrt{\omega^2-4T_c^2}$ (see
dashed line in Fig.~\ref{result}). Insight into this result can be gained
by analyzing the transport processes in the basis of the bonding
and antibonding states of the DQD,
$\ke{-}=\cos\theta\ke{L}+\sin\theta\ke{R}$,
$\ke{+}=-\sin\theta\ke{L}+\cos\theta\ke{R}$, respectively, with
$\tan\theta=2T_c/(\Delta+\epsilon)$ and with
$\epsilon=\sqrt{\Delta^2+4T_c^2}$ the distance in frequency
between the bonding and antibonding state. In this basis there are
two relevant physical processes, which lead to a change by one
phonon due to electron transport through the DQD. They consist of
the sequential occupation of the states $|0,n\rangle\to
|\pm,n\rangle\to |\mp,n\mp 1\rangle\to |0,n\mp 1\rangle$, as
illustrated in Fig.~\ref{processes}. Both processes are resonant
when the condition $\epsilon=\omega$ is satisfied, which
correspond to the dashed curve in Fig.~\ref{result}. The resonator
is hence cooled provided that the rate of the cooling process in
Fig.~\ref{processes}(a) is faster than the heating process in
Fig.~\ref{processes}(b), which is satisfied for $\Delta>0$.

A simple analytical result for the cooling rate $\gamma_0$ due to
electron-phonon coupling can be found in the limit of large $\Gamma_L$ and reads
\begin{widetext}
 \begin{eqnarray} \label{W0}
&&\gamma_0\simeq\frac{\alpha^2}{\omega^2}\left(\frac{4T_c^2\Gamma_R}{\Delta^2
+ 2 T_c^2 + \frac{\Gamma_R^2}{4}}\right) \frac{\omega^3\Delta
\pt{2T_c^2+\Gamma_R^2+\omega^2}}
{\omega^2\pt{\Delta^2+4T_c^2+\frac{5}{4}\Gamma_R^2-\omega^2}^2+\Gamma_R^2\pt{\Delta^2+2T_c^2+\frac{\Gamma_R^2}{4}-2\omega^2}^2}+O\pt{\frac{1}{\Gamma_L}}
\end{eqnarray} \end{widetext} showing that, for
$\omega\gg\Gamma_R$, it is maximum when $\epsilon\simeq \omega$.
Note that $\bar n_{0}$, in Eq.~(\ref{n}), reads \begin{eqnarray}
\bar n_{0}\simeq
\frac{\Gamma_R^2+4\pt{\Delta-\omega}^2}{16\Delta\omega}+O\pt{\frac{1}{\Gamma_L}}\label{n0}
\end{eqnarray} This equation has the same form found in sideband
cooling of trapped ions under specific conditions~\cite{Cooling,Zippilli}.
The same equation (for other physical parameters) was derived
in~\cite{Wilson-Rae,Marquardt,Kippenberg2,MartinPZ04}, where
cooling of nanomechanical resonators with photons was mapped to
sideband cooling of trapped ions (the detailed theory can be found
in~\cite{Genes}). In our case, small values $\bar n_0\ll 1$ are
achieved for $\omega\gg\Gamma_R$. The minimum value $\bar
n_0=\Gamma_R^2/16\omega^2$ is reached for $\Delta=\omega$,
independent of the value of $T_c$. However, $\gamma_0$ is not
maximum for $\Delta=\omega$ as for other
schemes~\cite{Wilson-Rae,Marquardt,Kippenberg2,MartinPZ04} and
optimal cooling is found as a compromise between maximizing
$\gamma_0$ and minimizing $\bar n_0$, see Eq.~(\ref{n}) and
Fig.~\ref{result}. In particular, for $\gamma_0\gg\gamma_p$, then
$\bar n_{\rm St}\simeq \bar n_0+\pt{\bar n_p-\bar
n_0}\gamma_p/\gamma_0$ and the resonator can be cooled to the
ground state.

\begin{figure}[!th] \begin{center}
\includegraphics[width=8cm]{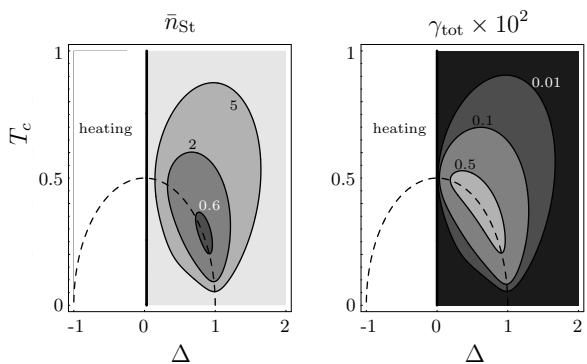} \caption{Same
as in Fig.~\protect\ref{result} with
$\Gamma_d=0.05\omega$.}\label{Figdeph}
 \end{center}
  \end{figure}

We now discuss how the above cooling efficiency is affected by
electronic dephasing inside the DQD (i.e., loss of coherence other
than incoherent tunneling from and to the electrodes). We model
this mechanism setting ${\mathcal
L}_d\rho=\Gamma_d\left(\sigma_z\rho\sigma_z-\rho\right)$ in
Eq.~(\ref{L_DD}), where $\sigma_z=|R\rangle\langle
R|-|L\rangle\langle L|$ and $\Gamma_d$ is the dephasing
rate~\cite{Viola}. We find that its effect on the cooling dynamics
is to lower the efficiency by broadening the mechanical
resonances. The resonator can be cooled when $\Gamma_d<\omega$ and
smaller values of $\Gamma_d$ give higher cooling efficiencies.
Figure~\ref{Figdeph} displays $\bar n_{\rm St}$ and $\gamma_{\rm
tot}$ for $\Gamma_d=0.05\omega$, while the other parameters are
the same as in Fig.~\ref{result}. For these parameters, the
resonator is cooled from $\bar n_p=24$ to $\bar n_{\rm St}\sim
0.5$ (corresponding to $T_{\rm osc}\sim 4$ mK for
$\omega=2\pi\times 100$ MHz).

We now summarize the relevant conditions for ground-state cooling.
The cooling mechanism is based on the resonant absorption of a
phonon by electron transport through the DQD. We have shown that
the resonance condition is found by tuning $\Delta$ and $T_c$ such
that $\omega=\sqrt{\Delta^2+4T_c^2}$, provided that $\Delta>0$.
Resonant enhancement of phonon absorption can be achieved when the
lifetimes of the electronic states inside the DQD are long as
compared to the period of the mechanical oscillations, i.e.,
$\Gamma_R,\Gamma_d\ll\omega$~\cite{footnote:4}. We remark that
this condition cannot be achieved for an individual quantum dot, where the resonance
is thermally broadened by the Fermi-Dirac distribution of the
electrons in the electrodes.

Most parameters of the proposed setup can be tuned. The frequency
$\omega$ depends on the length of the nanotube section that is
suspended, while the frequency difference $\Delta$, the tunneling
rates $T_c$, $\Gamma_L$, and $\Gamma_R$, can be controlled by the
external gates~\cite{Gotz}. However, little is known on electron
dephasing in nanotubes: the rate $\Gamma_d$ remains to be measured
and it is not clear whether $\Gamma_d$ can be tailored so to
fulfill $\Gamma_d<\omega$. Note that the proposed cooling scheme
can be applied to other device layouts. For instance, our
calculations can be employed for mechanical resonators
electrostatically coupled to fixed DQDs, microfabricated with
metal or semiconducting material~\cite{Naik}.

The effective temperature achieved by cooling could be probed by
measuring the amplitude of the oscillation fluctuations as a
function of the oscillator frequency. One possible strategy is to
monitor the microwave signal reflected off an external $LC$ tank
that is coupled to the DQD~\cite{Naik}. The signal might be
improved by switching the barriers along the nanotube to higher
transmissions once the phononic mode has been cooled down. There
would be a short time window ($<1/\gamma_p$) for measurements
before the effective temperature increases.

In conclusion we have shown that a high-Q flexural mode of a
carbon nanotube can be cooled to the ground state with constant
electron currents. The possibility to manipulate such macroscopic
objects using matter waves open novel avenues for quantum
manipulation of this kind of mesoscopic systems. An interesting
outlook is to use these concepts for realizing quantum reservoir engineering, in the
spirit of~\cite{Rabl04}, in order to achieve other kinds of
non-classical states.

We thank D. Garcia for discussions.
We acknowledge support by the European Commission (EURYI; EMALI
MRTN-CT-2006-035369; FP6-IST-021285-2), and by the Spanish
Ministerio de Educaci{\'o}n y Ciencia (QOIT; Consolider-Ingenio
2010; QNLP, FIS2007-66944; Ramon-y-Cajal; Juan-de-la-Cierva).


\begin{thebibliography}{99}
\bibitem{footnote:4} Note that $\Gamma_L$ does not affect
relevantly the linewidth of the electronic states $|L\rangle$ and
$|R\rangle$, since the current goes from left to right. This is
correct as long as $\Gamma_L$ is the largest rate.
\end{thebibliography}
\end{document}